\documentclass[11pt,a4paper]{article}
\begin{document}

\title{ Klein-Gordon equation and the stable problem in the Rindler space-time  \footnote{ E-mail of Tian:
 hua2007@126.com, tgh-2000@263.net}}
\author{Tian Gui-hua,\ \ Shi-kun Wang,  \ \ Shuquan Zhong\\
School of Science, Beijing University \\
of Posts And Telecommunications. Beijing100876, China.\\Academy of
Mathematics and Systems Science,\\ Chinese Academy of
Sciences,(CAS) Beijing, 100080, P.R. China.}
\date{}
\maketitle

\begin{abstract}
The Klein-Gordon equation in the Rindler space-time is studied
carefully. It is shown that the stable properties depend on using
what time coordinate to define the initial time. If we use the
Rindler time, the scalar field is stable. Alternatively, if we use
the Minkowski time, the scalar field may be regarded unstable to
some extent.   Furthermore, the complete extension of the Rindler
space time is the Minkowski space time, we could also study the
stable problem of the Rindler space time by the Klein-Gordon
equation completely in the Minkowski coordinates system. The
results support that the Rindler space time is really unstable.
This in turn might cast some lights on the stable problem of the
Schwarzschild black-hole, which not only in many aspects shares
the  similar geometrical properties with the Rindler space time
but also has the very same situation in stable study as that in
Rindler space time. So, it is not unreasonable to infer that the
Schwarzschild black hole might really be unstable in comparison
with the case in Rindler space time. Of course, one must go
further to get the conclusion definitely.

\end{abstract}

We have studied the stable problem in the Schwarzschild black hole
recently\cite{tian1}-\cite{tian4}. Usually, researchers gets the
conclusion that it is stable by using the Schwarzschild time $t=0$
to define the initial time\cite{2},\cite{wald}. It is recently
noticed that the Schwarzschild time $t=0$ could not be used to
study the stable problem due to the fact that $g_{00}|_{r=2m}=0$
makes the Schwarzschild time $t$ loses its meaning at the horizon
$r=2m$. By using the Kruskal time coordinate $T=C=const$, the
stable properties depend on the sign of the initial time
$T=C=const$: the Schwarzschild black hole is stable when $T=C \ge
0$, whereas the Schwarzschild black hole is unstable when $T=C \le
0$\cite{tian1}-\cite{tian4}. These unexpected results are in
contrast with the conclusion taken as granted that the
Schwarzschild black hole is stable. Which one is really correct?
There still is not a convincing and satisfying answer to it due to
the fact it is almost impossible to restudy the problem completely
in the Kruskal coordinates whose background metric is varying with
time\cite{2}.

The Rindler space time is one part of the Minkowski space time
whose constant spatial coordinates describe   accelerated
observers with constant accelerations. The interest in Rindler
space time lies in its similar geometrical structure with the
Schwarzschild black-hole. Actually, the Hawking radiation in the
Schwarzschild black-hole is closely connected with the Unruh
effect in the Rindler space time\cite{1},\cite{3}. Apart from the
similar geometrical properties with the Schwarzschild black-hole,
the Rindler space time is mathematically simple to study the
physical problem and has closed form for them.

Here we study the Klein-Gordon equation and the stable problem of
the Rindler space time, and try to find what effect the horizon of
the space time may have on its stable study. Subsequently we want
to obtain some clues to the stable study of the Schwarzschild
black hole which is similar very much with that of the Rindler
space time\cite{tian1}-\cite{tian4}.

\section*{I.The scalar field equation in the Rindler space time}
The metric of the Rindler space time is
\begin{equation}
ds^{2}=-z^2dt^{2}+dz^{2}+dx^{2}+ dy^{2}\label{orimetric-r}
\end{equation}
with $0<z<+\infty $, $t,\ x,\ y\in (-\infty,\ +\infty)$. The
Minkowski space time's metric is
\begin{equation}
ds^{2}=-dT^{2}+dZ^{2}+dx^{2}+ dy^{2},\label{orimetric-m}
\end{equation}
and the transformation equations between them are
\begin{equation}
z=\sqrt{Z^2-T^2}\label{relation zZ1}
\end{equation}
and
\begin{equation}
e^{2t}=\frac {Z+T}{Z-T},\label{relation zZ0}
\end{equation}
or
\begin{equation}
Z=z\cosh t\label{relation Zz1}
\end{equation}
and
\begin{equation}
T=z\sinh t.\label{relation Zz0}
\end{equation}
Of course, the Rindler space time corresponds the part $Z>0,\
Z^2-T^2>0$ of the Minkowski space time. $Z^2-T^2=0$ corresponds
the horizon of the Rindler space time. The horizon is denoted by
$z=0$ using the Rindler coordinates and the metric
(\ref{orimetric-r}) is singular at the horizon.

The Klein-Gorndon equation of the scalar field  is
\begin{equation}
\frac 1 {\sqrt{-g}} \frac {\partial }{\partial x^{\mu}}\left[\frac
1 {\sqrt{-g}}g^{\mu \nu}\frac {\partial \Psi}{\partial
x^{\nu}}\right]-m^2\Psi =0, \label{kg eqwuation}
\end{equation}
which turns out as
\begin{equation}
-\frac 1 {z}\frac{\partial^{2}\Psi}{\partial t^{2}}+\frac 1 {z}
\frac {\partial }{\partial z}\left[z\frac {\partial \Psi}{\partial
z}\right]+ \frac{\partial^{2}\Psi}{\partial
x^{2}}+\frac{\partial^{2}\Psi}{\partial
y^{2}}-m^{2}\Psi=0\label{kg in r}
\end{equation}
in the Rindler coordinates.

The coordinates $t,\ x,\ y$ range from $-\infty $ to $+\infty $,
therefore, the normal mode decomposition of the scalar field
$\Psi$ must be
\begin{equation}
\Psi =\psi(z) e^{-i\omega t+ik_1x+ik_2y}
\end{equation}
with $\psi$ satisfying the following equation
\begin{equation}
 \frac {d}{d z}\left[z\frac {d\psi}{d z}\right]-\frac {\omega
 ^2}z\psi+
z\left[m^{2}+k_1^2+k_2^2\right]\psi=0.\label{kg2 in r}
\end{equation}
The equation (\ref{kg2 in r}) is Bessel's equation with its order
being $\pm i\omega $\cite{1}. Its solutions are the modified
Bessel function $I_{\pm i\omega }(\sqrt{m^{2}+k_1^2+k_2^2}\
z)$.\footnote{when $i\omega $ is an positive integer, $I_{-
i\omega }$ is replaced by $K_{- i\omega }$.} Suppose the frequency
$\omega =\alpha +i\beta $ with $\alpha , \ \beta $ real and $\beta
>0$, then $I_{- i\omega }(\sqrt{m^{2}+k_1^2+k_2^2}\ z)$ goes to
infinity exponentially as $z\rightarrow \infty $ and vanishes at
$z=0$. On the contrary, $I_{+ i\omega }(\sqrt{m^{2}+k_1^2+k_2^2}\
z)$ goes to infinity at $z=0$ and falls off to zero exponentially
at $z\rightarrow \infty $\cite{1}.

The  boundaries of the  equation (\ref{kg2 in r}) consist of
$z=0$, the horizon $H^ \pm $ of the Rindler space time, and
$z\rightarrow \infty $ of the infinity. We must demand the scalar
field function $\Psi$ initially well-behaved at the boundaries.
For the infinity, we could require the field $\Psi$ falling off to
zero initially. So we just select $I_{i\omega
}(\sqrt{m^{2}+k_1^2+k_2^2}\ z)$ to satisfying the infinity
boundary condition initially with the frequency whose imaginary is
positive:
\begin{equation}
\Psi = e^{-i\omega t+ik_1x+ik_2y}I_{ i\omega
}(\sqrt{m^{2}+k_1^2+k_2^2}\ z)\label{I1 in rindler}.
\end{equation}

But the metric is singular at the horizon $z=0$, and the time
coordinate $t$ even loses its meaning there due to the fact
$g_{00}|_{z=0}=0$. Here comes the problem to choose what time
coordinates to define the initial time. The same problem appears
in the stability study of the Schwarzschild black
hole\cite{tian1}-\cite{tian4}. The problem really results in many
controversial conclusions regarding whether or not  the
Schwarzschild black hole is stable\cite{tian1}-\cite{tian4} and
still is unsolved. Just as in the case of stable study in the
Schwarzschild black hole, there are obviously two kinds time
coordinates to use, that is, the Rindler time $t$ and the
Minkowski time $T$. In the following, we have these two choices
comparably and carefully compare the corresponding results.

\subsection*{I.1 The initial time is defined by the Rindler time $t=0$}

By eq.(\ref{I1 in rindler}), the scalar field is regular at the
infinity at $t=0$. But  $I_{ i\omega }(\sqrt{m^{2}+k_1^2+k_2^2}\
z)$ blows up at the horizon $z=0$ at $t=0$, that is, $\Psi $ is
initially blows up at the horizon for the positive imaginary of
the frequency  ($\beta >0$). This shows that $\beta >0$ is not
accepted for the scalar field because it behaves badly at the
initial time $t=0$. Therefore the scalar field is stable and the
Rindler space time is also stable to some extent ( concerning to
the perturbation of the scalar field ).

\subsection*{I.2 The initial time is defined by the Minkowski time
$T$}

We now use the Minkowski time $T$  to define the initial time. By
the transformation equations (\ref{relation zZ1}),(\ref{relation
zZ0}), we could rewrite the scalar field $\Psi$ as the function of
the independent variables $T,\ Z,\ x,\ y$:
\begin{equation}
\Psi =A e^{\frac{-i\omega
t}2\ln{\frac{Z+T}{Z-T}}}e^{ik_1x+ik_2y}I_{ i\omega
}(\sqrt{m^{2}+k_1^2+k_2^2}\sqrt{Z^2-T^2})\label{I2 in Minkowski}.
\end{equation}
Excluding the horizon, the scalar field $\Psi$ is bounded at the
initial time $T=C=const$ for the positive frequency $\Im \omega
=\beta >0$. At the horizon $Z^2-T^2=0$, the asymptotic form of the
scalar field $\Psi$ in eq.(\ref{I2 in Minkowski}) is
\begin{equation}
\Psi \propto A \left({\frac{Z+T}{Z-T}}\right)^{\frac{-i\omega
}2}\left(Z^2-T^2\right)^{\frac{i\omega }2}=A
\left(Z-T\right)^{i\omega } \label{I3 in Minkowski}.
\end{equation}
Whether or not the equation (\ref{I3 in Minkowski}) is divergent
initially at the horizon depends really on the sign of the initial
time $T=C$. When we select the initial time $T=C\ge 0$, the
horizon initially corresponds $Z=T=C$. So, the scalar field $\Psi$
goes to infinity initially at $Z=T=C$. This excludes the
possibility of the positive imaginary frequency, and subsequently
ensures its stability and the Rindler space time respectively to
some extent.

Alternatively, we could also select the initial time $T=C\le 0$.
The horizon initially corresponds $Z=-T=-C$, so the scalar field
$\Psi$ is also bounded initially at the horizon for the positive
imaginary frequency $\beta >0$. In this very case, the scalar
field $\Psi$ is not stable and the Rindler space time consequently
is also unstable with respect to the scalar field perturbation.

\subsection*{I.3 Discussion}
We use different initial times to study the scalar field equation
in Rindler space time, and obtain the completely controversial
conclusions on its stable study. Which one is really corresponds
real or correct one? If we only study the scalar field equation
completely in the Rindler space time, the selection is difficult
to make. Actually, the same situation appears also in the stable
study of the Schwarzschild black hole, and the same controversial
conclusions are obtained there\cite{tian1}-\cite{tian4}. This kind
problem is unsolved in the Schwarzschild black hole. It is
suggested one should study the stable problem of the Schwarzschild
black hole completely in the Kruskal coordinate system to get a
definite answer to the controversy\cite{2}. But it becomes almost
impossible because the background metric in the Kruskal coordinate
system is not even stationary. So, one could not get absolutely
and convincingly definite answer to the stable problem of the
Schwarzschild black hole.

The geometry of the Rindler space time is much similar with that
of the Schwarzschild black hole, for example, the horizon, the
maximum extension, etc. Nevertheless, the extension of the Rindler
space time is the Minkowski space time, which is the simplest
space time and is static in contrast to the Kruskal space time.
Therefore, we could study the stable problem of the scalar field
in Rindler space time completely by the Minkowski space time
coordinate system. Subsequently, we could obtain definite answer
to it. This in turn might give some clues to the stable problem of
the Schwarzschild black hole.

\section*{II. The scalar field equation in the Minkowski space time}

The scalar field equation in the Minkowski space time is
\begin{equation}
-\frac{\partial^{2}\Psi}{\partial T^{2}}+ \frac {\partial ^2
\Psi}{\partial Z^2}+ \frac{\partial^{2}\Psi}{\partial
x^{2}}+\frac{\partial^{2}\Psi}{\partial
y^{2}}-m^{2}\Psi=0.\label{kg in m}
\end{equation}
The Rindler space time corresponds the region where the
coordinates $T,\ x,\ y$ range from $-\infty $ to $+\infty $ and
$Z>0,\ Z^2-T^2>0$, therefore, the normal mode decomposition of the
scalar field $\Psi$ must be
\begin{equation}
\Psi = Ae^{-i\omega T+ik_3Z+ik_1x+ik_2y}
\end{equation}
where $k_3$ must satisfy the following equation
\begin{equation}
k_3=\pm \sqrt{\omega^2-k_1^2-k_2^2-m^2}.
\end{equation}
In order to make the scalar field $\Psi$ initially finite over
$-\infty<x,\ y<+\infty $, the numbers $k_1,\ k_2$ of course are
real. If we similarly choose the frequency $\omega =i\beta $ with
$\beta
>0$, the above equation becomes
\begin{equation}
k_3=\pm i\sqrt{\omega^2+k_1^2+k_2^2+m^2}.
\end{equation}
We could select $k_3=i\gamma $ with
$\gamma=\sqrt{\omega^2+k_1^2+k_2^2+m^2}>0$, then the scalar field
$\Psi$ is
\begin{equation}
\Psi =Ae^{\beta T}e^{-\gamma Z+ik_1x+ik_2y}.\label{psi in M}
\end{equation}
The boundaries of the Rindler space time consist of the infinity
$Z\rightarrow +\infty$ and the horizon $Z^2-T^2=0$. Because the
Minkowski metric is regular everywhere including the horizon, we
could directly use the Minkowski time $T=0$ to define the initial
time. By eq.(\ref{psi in M}), we could see the scalar field $\Psi$
is well-behaved everywhere corresponding to the Rindler space time
at the initial time $T=0$. But it blows up to infinity as the time
$T$ goes to infinity for $\beta >0$.

Therefore, the scalar field is unstable in the Rindler space time,
and consequently the Rindler space time is unstable with respect
to the scalar field perturbation. The conclusive instability of
the Rindler space time also tells us which one is correct
concerning the study of the problem in the Rindler space time in
section I. It definitely support the view that the Rindler time
coordinate $t$ is not suitable to use to study the stable problem
containing the horizon where $t$ loses its meaning. It also
conclusively shows  only the instability in section I by using the
Minkowski time $T=C<0$ is correct. This might give some clues
concerning the stable problem of the Schwarzschild black hole.

Though the Kruskal metric of the Schwarzschild black hole is not
stationary, and one has not obtained the same conclusive result as
that in the Minkowski metric of the  Rindler space time; it is
nevertheless not unreasonable to infer that the Schwarzschild
black hole might really be unstable in comparison with the case in
Rindler space time. Of course, one must go further to get the
conclusion definitely.

\section*{Acknowledgments}

We are supported in part by the national Natural Science
Foundations of China under Grant  No.10475013, No.10373003,
No.10375087, the National Key Basic Research and Development
Programme of China under Grant No 2004CB318000
 and the post-doctor foundation of
China

\end{document}